\documentstyle[11pt]{article}
\begin{document}
\thispagestyle{empty}

\begin{center}
               RUSSIAN GRAVITATIONAL ASSOCIATION\\
               CENTER FOR SURFACE AND VACUUM RESEARCH\\
               DEPARTMENT OF FUNDAMENTAL INTERACTIONS AND METROLOGY\\
\end{center}
\vskip 4ex
\begin{flushright}                              RGA-CSVR-006/94\\
                                                gr-qc/9406030
\end{flushright}
\vskip 45mm

\begin{center}
{\bf INTEGRABLE WEYL GEOMETRY IN MULTIDIMENSIONAL \\
     COSMOLOGY. NUMERICAL INVESTIGATION}

\vskip 5mm
{\bf M. Yu. Konstantinov, V. N. Melnikov }\\
\vskip 5mm
     {\em Centre for Surface and Vacuum Research,\\
     8 Kravchenko str., Moscow, 117331, Russia}\\
     e-mail: mel@cvsi.uucp.free.msk.su\\
\bigskip

\begin{abstract}
The evolution of 4-dimensional and (4+d)-dimensional (d=1,2)
cosmological models based on the integrable Weyl geometry are
considered numerically both for empty space-time and for scalar field
with non minimal coupling with gravity.
\end{abstract}

\vskip 50mm

             Moscow 1994
\end{center}
\pagebreak

\setcounter{page}{1}

\begin{center}
{\bf
INTEGRABLE WEYL GEOMETRY IN MULTIDIMENSIONAL \\
COSMOLOGY. NUMERICAL INVESTIGATION}

\vglue 0.4cm

{\tenrm  M. Yu. Konstantinov, V. N. Melnikov}
\vskip3mm
{\it Center for Surface and Vacuum Research,}\\
\baselineskip=11pt
{\it 8 Kravchenko str., Moscow, 117331, Russia}\\
{E-mail: mel@cvsi.uucp.free.msk.su}
\end{center}

\begin{abstract}
The evolution of 4-dimensional and (4+d)-dimensional (d=1,2)
cosmological models based on the integrable Weyl geometry are
considered numerically both for empty space-time and for scalar field
with non minimal coupling with gravity.
\end{abstract}

\section{ Introduction}
 The multidimensional gravitation theories are very attractive in the
context of the unification of fundamental interactions. Moreover,
several modern theories require space-time to have more than four
dimensions [1-8]. The nonobservability of additional dimensions in
such theories needs an explanation. Among different possible ways of
such explanation the hypothesis about dynamical contraction of
internal manifold during expansion of the universe is very popular.
This idea is realized in many exact cosmological solution of
multidimensional Einstein's equations [9-22]. As a rule such models
require additional fields and do not avoid initial big bang
singularity. The introduction of additional fields in
multidimensional gravitation theories destroy their pure geometrical
character and require an additional motivation [4]. Such motivation
may be done in the framework of some generalizations of Riemannian
geometry. In four dimensional case such generalization in several
cases leads to removing of cosmological big bang singularity [23-25].
That is why the unification of generalized geometric structures and
multidimensional gravity seems to be very attractive. Unfortunately,
only in several papers the multidimensional gravitation theory and
cosmology are considered in the scope of some generalization of
Riemannian geometry [26-28].

One of the simplest generalization of the Riemannian geometry is the
integrable Weyl geometry with the connection components
\begin{equation}   
\Gamma _{\beta \gamma }^{\alpha} =
\widetilde{\Gamma}^{\alpha}_{\beta \gamma } - \frac{1}{2}
\left( \omega_{\beta} \delta_{\gamma}^{\alpha} +\omega_{\gamma}
\delta_{\beta}^{\alpha}-g_{\beta \gamma }\omega^{\alpha} \right),
\end{equation}
where $\widetilde{\Gamma}^{\alpha }_{\beta \gamma }$ are the
Christoffel symbols, $\omega_{\alpha} =\omega,_{\alpha} $, $\omega $
is a scalar field, $\delta^{\alpha}_{\beta}$ are the Kroneker
symbols, $g_{\alpha \beta }$ is a metric tensor; the small Greek
indices take values from $0$ to $n-1$, $n$ is a dimension of
space-time. The Ricci tensor and the curvature scalar of the
connection (1) are equal to
\begin{equation}         
R_{\mu\nu}=\widetilde{R}_{\mu \nu}+\frac{n-2}{2}\omega_{\mu\Vert\nu}+
 \frac{1}{2} g_{\mu \nu} \widetilde{\Box }\omega +
\frac{n-2}{4}\left( \omega_{\mu}\omega_{\nu} -
g_{\mu\nu}\omega^{\lambda}\omega_{\lambda} \right),
\end{equation}
\begin{equation}         
 R=\widetilde{R}+(n-1)\widetilde{\Box}\omega -
\frac{(n-1)(n-2)}4\omega^{\lambda} \omega _{\lambda},
\end{equation}
where the tildes denote the quantities calculated in the connection
$\widetilde{\Gamma }_{\beta \gamma}^\alpha $, two parallel
vertical bars and $\widetilde{\Box }$ denote the covariant derivative
and the d'Alembert operator of this connection. It is necessary to
note that the integrable Weyl space-time is also
conformally-Riemannian, since there is a conformal transformation of
metric tensor $g_{\alpha \beta }$ which maps the Riemannian
space-time into integrable Weyl space-time. As the integrable Weyl
space-time is defined by the pair $\left(g_{\alpha \beta }, \omega
\right)$ the gravitation theory in this space-time does not coincide
with Einsteinian general relativity because the field $\omega$ must
be contained in the Lagrangian independently from $g_{\alpha \beta }$
and cannot be excluded by the conformal transformation.

Some features of the Einsteinian cosmological models with scalar
fields were recently considered by several authors
[9,11,19,21,28,29-37] both in 4-dimensional and in (4+d)-dimensional
space-times. The cosmological models in four-dimensional
Weyl-integrable space-time were recently considered by Novello et al.
in [23], where the existence of nonsingular open cosmological models
was shown. The appearance of Weyl geometry in multidimensional
cosmology was discussed also in [27].

In this paper we consider the influence of Weyl geometry on the
evolution of Friedman-Robertson-Walker (FRW) cosmological models in
multidimensional gravitation theory. As usually the space-time is
assumed to have the structure of direct product $M^{4} \times V^{d}$
of four-dimensional FRW space-time $M^{4}$ and $d$-dimensional
interior space $V^{d}$ that is supposed to be d-sphere $S^{d}$ or
$d$-torus $T^{d}$. The metric of space-time is supposed to be
block-diagonal
\begin{equation}                    
ds^{2}=dt^{2}-a^{2}(t) \left( \frac{dr^{2}}{1-kr^{2}} +
r^{2}d \Omega^{2} \right) -\widetilde{g}_{ab}du^{a}du^{b},
\end{equation}
where $k=+1$, $0$, $-1$ for closed, plane and open models,
$d\Omega^{2}$ is a line element on two-sphere, $u^{a}$, $a=1,...,d$,
and $wide\tilde{g}_{\alpha \beta }$ are the coordinates and metric
tensor of the interior space $V^{d}$. Once we consider only spatially
homogeneous FRW cosmologies, it is natural to make the Weyl scalar
field $\omega$ to be a function of cosmic time $t$ only:
$\omega=\omega(t)$. We consider both vacuum case and non vacuum case
with the additional scalar field $\varphi$ with non minimal coupling.
The 4-dimensional case will be briefly considered also for
completeness.  The existence of the conformal map between Riemannian
and integrable Weyl space-times may be used for generation of exact
solutions from the known solutions of general relativity. Such
approach admits obtaining only the particular solutions.  Therefore
to demonstrate general qualitative behavior of the models we solve
the system of cosmological equations numerically with initial values
given at $t=0$ and satisfying the constraint equation.  For that
purpose we use adaptive numerical methods with automatic choice of
integration step and with the stiffness checking.  The geometrical
units where $G=c=1$ are used in what follows.

\section{Integrable Weyl cosmology in vacuum}

Following [23] we shell consider the vacuum cosmological models in
the gravitation theory with the Lagrangian
\begin{equation}        
L=R+\xi \omega _{\alpha} \omega ^{\alpha}
\end{equation}
where R is defined by (3) and $\xi=const$. After excluding
the total derivatives of the scalar field Lagrangian (5) takes the
form
\begin{equation}        
L=\widetilde{R} -
\frac{(n-1)(n-2)-4\xi }{4} \omega^{\alpha} \omega_{\alpha}
\end{equation}
So, the theory differs from the Einstein theory with the massless
scalar field by the coefficient before the square of the scalar field
gradient and has different geodesic lines.  Note also that due to the
definition of the Weyl connection (1) the scalar field $\omega$
cannot be renormalized and hence the coefficient $\xi$ before
$\omega^{\alpha}\omega_{\alpha}$ cannot be put to $\pm 1$ as it may
be done in the pure Einstein theory with massless scalar field.
Variation of (6) with respect to the pair
$\left( g_{\alpha\beta}, \omega \right)$ of independent variables
yields the equations
\begin{equation}        
\widetilde{R}_{\mu \nu} - \frac{1}{2} g_{\mu \nu } \widetilde{R}-
\frac{(n-1)(n-2)-4\xi }{4} \left( \omega _{\mu} \omega _{\nu} -
\frac{1}{2}g_{\mu \nu }\omega^{\alpha} \omega _{\alpha} \right) =0,
\end{equation}
and
\begin{equation}        
 \widetilde{\Box }\omega =0
\end{equation}
The equations (7), (8) coincide with the Einstein equations for the
massless scalar field, whose solutions for the FRW cosmological
models were investigated both in four-dimensional [32] and
multidimensional cases [37]. By this reason here we only summarize
briefly the main results.

{\it 2.1 Four-dimensional case}. As the scalar field $\omega$ is a
function on t only, equation (8) yields the first integral
\begin{equation}        
\dot{\omega} = \frac{\gamma}{a^{3}}
\end{equation}
where overdot denotes time differentiation and $\gamma=const$ is
the integration constant.  Due to (9) equations (7) take the form
\begin{equation}           
  \dot{a}^{2} + k - \frac{\lambda \gamma^{2}}{12a^{4}}=0
\end{equation}
and
\begin{equation}             
    2a\ddot{a}+\dot{a}^{2}+k+\frac{\lambda \gamma^{2}}{4a^{4}}=0
\end{equation}
where $\lambda=(3-2\xi)$. As it is easy to see from (10), only
singular and static solution of equations (10)-(11) exist if
$\lambda>0$. For negative values of $\lambda$ solution exists only
for the open models. In this case
$a(t) \geq a_{0} = (\xi-3)\gamma^{2}/12$ and so the cosmological
singularity is absent. The qualitative behavior of scale factor
$a(t)$ for negative $\lambda$ is shown at figure 1 and its features
are discussed in detail in [23].

\medskip

\begin{center}
 Fig. 1. \\
Qualitative behavior of the scale factor $a(t)$ of the open universe
in four-dimensional Weyl-integrable space-time.
\end{center}

{\it 2.2. Multidimensional case}. In the multidimensional case the
behavior of the model depends not only on the parameter $\xi$, as in
the previous case, but on the structure of the interior space also.
For simplicity only 5- and 6-dimensional models will be considered in
the following. We consider these two cases separately. The main
qualitative features of models in general $n$-dimensional ($n>6$)
case are the same as in 5- and 6-dimensions.

{\it 2.2.1. 5-dimensional models}. In 5-dimensions space-time
interval (4) reads
\begin{equation}             
 ds^{2}=dt^{2}-a^{2}(t)\left( \frac{dr^{2}}{1-kr^{2}} +
r^{2}d\Omega^{2} \right) - s^{2}(t)du^{2}
\end{equation}
where $u \in S^{1}$ is the interior space coordinate. Assuming, as
above, a scalar field $\omega$ to be a function of the cosmological
time only, the first integral of equation (8) takes the form
\begin{equation}             
  \dot{\omega} = \frac{\gamma_{1}}{a^{3}(t)s(t)},
\end{equation}
 where $\gamma_{1}=const$. Due to (12)-(13) equations (7) become
after simplification
\begin{equation}               
 3\frac{\dot{a}}{a}\frac{\dot{s}}{s} +
3\left( \frac{\dot{a}}{a} \right)^{2}+ \frac{3k}{a^{2}} -
    \frac{\gamma_{1}^{2}(3-\xi)}{2a^{6}s^{2}} = 0,
\end{equation}
\begin{equation}              
 \frac{\ddot{a}}{a}+2\left( \frac{\dot{a}}{a} \right)^{2} +
     \frac{\dot{a}}{a}\frac{\dot{s}}{s} +\frac{k}{a^{2}}=0,
\end{equation}
and
\begin{equation}              
 \frac{\ddot{s}}{s} + 3\frac{\dot{a}}{a} \frac{\dot{s}}{s}=0
\end{equation}
The last equation has the first integral
\begin{equation}              
  \dot{s} =\frac{\gamma_{2}}{a^{3}}
\end{equation}
where $\gamma_{2}=const$. It is easy to see that analogous to the
four-dimensional case the nonsingular solutions of equations (15),
(17) exist only for the open models ($k=-1$). In this case for $t<0$
the scale factor of 3-space $a(t)$ decreases monotonically from
infinity to its minimal value $a_{0}$ and then grows to infinity at
$t>0$, while the Weyl field $\omega(t)$ and the scale factor of
interior space evaluates monotonically from
$\omega_{-}= \lim_{t \rightarrow -\infty} \omega(t)$ and
$s_{-}= \lim_{t \rightarrow -\infty} s(t)$ to
$s_{+}=\lim_{t\rightarrow \infty} s(t)$, where $a_{0}$,
$\omega_{\pm}$ and $s_{\pm}$ are defined by the integration constants
and may have arbitrary values.  Note that if $\gamma_{2}<0$ than the
constants $s_{-}$ and $s_{+}$ satisfy the condition $s_{-}>s_{+}$ and
so the standard dimensional reduction scenario is realized. The
typical shape of the functions $a(t)$ and $s(t)$ are shown on the
figures (2.a,b).

\medskip

\begin{center}
 Fig. 2. \\
The qualitative behavior of the scale factors $a(t)$ (2a) and $s(t)$
(2b) of five-dimensional Weyl integrable cosmological model with
open 3-space.
\end{center}

The figure (2a) shows that unlike the 4-dimensional case the
evolution of 3-space in 5-dimensional model is time-asymmetric. This
asymmetry appears because the equation (15) depends not only on
$\dot{s}(t)$ but also on the time-asymmetric interior space scale
factor $s(t)$.

{\it 2.2.2. 6-dimensional models}. In 6-dimensional case we
consider two types of topological structures for the interior space:
the 2-sphere $S^{2}$ and 2-dimensional torus $T^{2}$. Therefore the
space-time metric (4) may have one of two forms
\begin{equation}         
ds^{2}=dt^{2}-a^{2}(t) \left( \frac{dr^{2}}{1-kr^{2}} +
 r^{2}d\Omega^{2} \right) - s^{2}(t) \left( \frac{du^{2}}{1-u^{2}} +
u^{2}dv^{2} \right)
\end{equation}
or
\begin{equation}         
   ds^{2}=dt^{2}-a^{2}(t) \left( \frac{dr^{2}}{1-kr^{2}} +
r^{2}d\Omega^{2} \right) - s_{1}^{2}(t)du^{2}-s_{2}^{2}(t)dv^{2} ,
\end{equation}
where $\{ u, v \}$ are the coordinates on $S^{2}$ or $T^{2}$
respectively. First integrals of equation (8) take the form
\begin{equation}             
   \dot{\omega} = \frac{q_{1}}{a^{3} s^{2}},
\end{equation}
for metric (18) and
\begin{equation}             
   \dot{\omega} = \frac{q_{2}}{a^{3}s_{1}s{2}},
\end{equation}
for metric (18).

Equations (7) for the metric (18) after simplification take the form
\begin{equation}             
 \frac{\ddot{a}}{a}+2 \left( \frac{\dot{a}}{a} \right)^{2} +
   2 \frac{\dot{a}}{a}\frac{\dot{s}}{s} + \frac{2k}{a^{2}} = 0,
\end{equation}
\begin{equation}             
 \frac{\ddot{s}}{s} + \left( \frac{\dot{s}}{s} \right)^{2} +
   3 \frac{\dot{a}}{a}\frac{\dot{s}}{s} + \frac{1}{s^{2}} = 0,
\end{equation}
and the constraint equation
\begin{equation}             
 3\frac{\dot{a}}{a} \left(\frac{\dot{a}}{a} +
2\frac{\dot{s}}{s}\right) + \frac{3k}{a^{2}} + \frac{1}{s^{2}} -
\frac{q_{1} \left(5-\xi \right)}{2a^{6}s^{4}}=0.
\end{equation}
The first two equations are dynamical and the last is the constraint.

It is easy to see that only singular solutions of equations (22)-(24)
exist: the scale factor $s(t)$ of the interior space evolves from zero
at $t=t_{0}$ to its maximal value $s_{max}$ and return to zero at
$t=t_{1}>t_{0}$. The behavior of $a(t)$ depends on the sign of $k$.
Namely, if $k=+1$ then the qualitative evolution of $a(t)$ is the
same as the evolution of $s(t)$. If $k=0$ than $a(t)$ increase from
zero at $t=t_{0}$ to infinity at $t=t_{1}$ or decrease from infinity
to zero; the unstable solutions with $a(t)=c0nst$ are also exist.
Finally, if $k=-1$ then $a(t)$ evolves from infinity at $t=t_{0}$
to its minimum $a_{min}$ and then grows to infinity at $t=t_{1}$.

For the metric (19) equations (7) after simplification read
\begin{equation}           
 \frac{\ddot{a}}{a}+
\frac{\dot{a}}{a} \left(\frac{\dot{s}_{1}}{s_{1}} +
\frac{\dot{s}_{2}}{s_{2}} \right) +
2 \left(\frac{\dot{a}}{a} \right)^{2}+\frac{2k}{a^{2}} = 0,
\end{equation}
\begin{equation}           
 \frac{\ddot{s}_{1}}{s_{1}} +
3 \frac{\dot{a}}{a}\frac{\dot{s}_{1}}{s_{1}} +
 \frac{\dot{s}_{1}}{s_{1}}\frac{\dot{s}_{2}}{s_{2}} = 0,
\end{equation}
\begin{equation}           
 \frac{\ddot{s}_{2}}{s_{2}} +
 3\frac{\dot{a}}{a}\frac{\dot{s}_{2}}{s_{2}} +
 \frac{\dot{s}_{1}}{s_{1}}\frac{\dot{s}_{2}}{s_{2}} = 0,
\end{equation}
and the constraint equation
\begin{equation}            
 3\frac{\dot{a}}{a} \left( \frac{\dot{a}}{a} +
 \frac{\dot{s}_{1}}{s_{1}} + \frac{\dot{s}_{2}}{s_{2}} \right) +
 \frac{\dot{s}_{1}}{s_{1}}\frac{\dot{s}_{2}}{s_{2}} +
 \frac{3k}{a^{2}} -
\frac{q_{2}^{2} \left(5-\xi \right)}{2a^{6}s_{1}^{2}s_{2}^{2}} = 0.
\end{equation}

As in 4- and 5-dimensional cases the nonsingular solutions of the
equations (25)-(28) exist only for the open models ($k=-1$).
Analogously to 5-dimensional case the scale factor of 3-space $a(t)$
in these models decreases monotonously from infinity to its minimal
value $a_{0}$ and then grows to infinity at $t \rightarrow +\infty$,
while the scale factors $s_{i}(t)$, $i=1$, $2$, of interior space
changes monotonously from $s_{i-}=\lim_{t \rightarrow -\infty}
s_{i}(t)$ to $s_{i+}=\lim_{t \rightarrow \infty} s_{i}(t)$. The
necessary condition for the realization of the dimensional reduction
scenario in this case are defined by the following inequalities
\begin{equation}            
    \frac{3}{a_{0}^{2}}+
  \frac{q_{2}^{2}(5-\xi)}{2a_{0}^{6}s_{1 0}s_{2 0}} > 0,
\end{equation}
and
\begin{eqnarray}            
  \dot{s}_{1}(0) < 0,       \dot{s}_{2}(0) < 0
\end{eqnarray}
It is necessary to note that inequality (29) is the necessary
condition for $\dot{s}_{1}$ and $\dot{s}_{2}$ to be of the same sign.
The time behavior of scale factors $a(t)$, $s_{1}(t)$ and $s_{2}(t)$
in this case is qualitatively the same as in 5-dimensional case
(Figure 2).

\section{Integrable Weyl cosmology in theory with non minimal scalar
field}

In this section we consider cosmological models in gravitation
theories with Lagrangian
\begin{equation}            
L=R \left( 1+\frac{1}{2(n-1)}\varphi^{2} \right) +
  \xi \omega^{\alpha} \omega _{\alpha} +
  \eta \varphi ^{\alpha} \varphi_{\alpha},
\end{equation}
where R is defined by (3), $\varphi$ is a real scalar field,
$\eta=\pm 1$ and $\xi=const$ as above. In the limiting case
$\varphi=const$ Lagrangian (31) coincides with (5) while in another
limiting case $\omega=const$ it coincides with the Lagrangian for the
conformal-invariant scalar field.

The substitution of (3) into (31) gives after simplification
\begin{equation}            
L=\widetilde{R} \left( 1+\frac{\varphi^{2}}{2(n-1)} \right) -
  \varphi \varphi^{\alpha} \omega _{\alpha} -
  \frac{(n-1)(n-2)-4\xi}{4} \omega^{\alpha} \omega_{\alpha} -
  \frac{(n-2)}{8}\varphi^{2} \omega^{\alpha} \omega_{\alpha} +
  \eta \varphi ^{\alpha} \varphi_{\alpha},
\end{equation}
where the total derivatives of the scalar fields are omitted.

Variation of (32) with respect to independent variables $g_{\mu\nu}$,
$\omega$ and $\varphi$ yields the equations
\begin{eqnarray}            
\left( \widetilde{R}_{\mu \nu } - \frac{1}{2}g_{\mu \nu}
 \widetilde{R} \right) \left( 1+\frac{\varphi ^{2}}{2(n-1)} \right)-
  \frac{(n-1)(n-2)-4\xi }{4} \left( \omega_{\mu} \omega_{\nu} -
\frac{1}{2}g_{\mu \nu } \omega ^{\alpha} \omega _{\alpha} \right)-
   \nonumber  \\
 \frac{1}{2}\varphi \left( \varphi,_{\mu} \omega,_{\nu} +
\varphi ,_{\nu} \omega,_{\mu} \right) -
\frac{n-2}{8}\varphi ^{2} \left( \omega,_{\mu} \omega ,_{\nu} -
\frac{1}{2}g_{\mu \nu }\omega ^{\alpha} \omega _{\alpha} \right) +
\frac{\varphi}{n-1} \left( g_{\mu \nu }\widetilde{\Box }\varphi -
\varphi ,_{\mu \Vert \nu } \right) + \nonumber  \\
 \varphi ,_{\mu} \varphi ,_{\nu} \left( \eta -\frac{1}{n-1} \right) +
g_{\mu \nu }\varphi ^{\alpha} \varphi _{\alpha} \left( \frac{1}{n-1} -
\frac{\eta}{2} \right) = 0,      \\
\left(\frac{(n-1)(n-2)-4\xi}{2} + \frac{n-2}{4} \varphi^{2} \right)
\widetilde{\Box}\omega - \varphi \widetilde{\Box} \varphi -
\varphi,_{\nu} \varphi,^{nu} = 0,
\end{eqnarray}
and
\begin{equation}             
 \eta \widetilde{\Box}\varphi - \left(\widetilde{\Box}\omega +
\frac{1}{n-1}\widetilde{R} -
\frac{n-2}{4}\omega,^{\nu}\omega,_{\nu} \right)\varphi = 0,
\end{equation}

Equation (35) shows that non-Riemannian nature of space-time geometry
in the considered model leads to the effective mass generation for
the scalar field $\varphi$.

{\it 3.1. Four-dimensional models}. In four-dimensional case the
equations (33)-(35) consist of the constraint equation
\begin{equation}            
 \left( \frac{\dot{a}}{a} + \frac{k}{a^{2}} \right)^{2} \left( 3 +
\frac{\varphi^{2}}{2} \right) +
\frac{\dot{a}}{a} \varphi \dot{\varphi} +
\frac{\eta}{2} \dot{\varphi}^{2} -
\frac{1}{2} \varphi \dot{\varphi} \dot{\omega} -
\frac{1}{8} \varphi^{2} \dot{\omega}^{2} -
\frac{3-2\xi}{4}  \dot{\omega}^{2} = 0,
\end{equation}
and three dynamical equations
\begin{equation}            
 \left( 2+\frac{\varphi^{2}}{3} \right) \frac{\ddot{a}}{a} +
\frac{1}{3}\varphi\ddot{\varphi} +
2\left( \frac{\dot{a}}{a} \right)^{2}
\left( 2+\frac{\varphi^{2}}{3} \right) + \frac{5}{3}
\frac{\dot{a}}{a} \varphi \dot{\varphi} + \frac{1}{3}
\dot{\varphi}^{2} + \left(2+\frac{\varphi^{2}}{3} \right)
\frac{2k}{a^{2}} = 0,
\end{equation}
\begin{equation}            
 \left( \frac{\varphi^{2}}{2}-2\xi+3 \right)\ddot{\omega} - \varphi
\ddot{\varphi} + \left( 9-6\xi+\frac{3}{2}\varphi^{2} \right)
\frac{\dot{a}}{a} \dot{\omega} - 3 \frac{\dot{a}}{a} \varphi
\dot{\varphi} - \dot{\varphi}^{2} = 0,
\end{equation}
and
\begin{equation}          
\eta \ddot{\varphi} - \varphi \ddot{\omega} +
   2 \varphi \frac{\ddot{a}}{a} +
   3 \frac{\dot{a}}{a} \left( \eta\dot{\varphi} -
   \varphi \dot{\omega} \right) +
   \frac{1}{2} \varphi \dot{\omega}^{2} + 2 \varphi \left(
   \frac{\dot{a}}{a} \right)^{2} + \frac{2k}{a^{2}} \varphi = 0.
\end{equation}

The coefficients before $\ddot{a}/a$, $\ddot{\omega}$ and
$\ddot{\varphi}$ in the equations (37)-(39) depend both on the
parameters $\xi$, $\eta$ and on the scalar field $\varphi$. The
determinant of the matrix of coefficients before $\ddot{a}/a$,
$\ddot{\omega}$ and $\ddot{\varphi}$ is equal to

\centerline{
$ d = \left( \frac{\eta}{2} - \frac{2}{3} \right) \varphi^{4} +
\left( 2\eta + \frac{4\xi}{3} - \frac{2\eta \xi}{3} -
4 \right) \varphi^{2} + 6\eta - 4 \eta \xi $.}

The points where $d=0$ are the singular points of the system
(37)-(39). These points are not described by the system (36)-(39)
because for fixed $\eta$ and $\xi$ equation $d=const$ defines not
more than four fixed values of $\varphi$ and the system (36)-(39)
reduces to the first order  system. Therefore the initial value of
the field $\varphi$ must be from the open set $d \neq 0$.

For $\eta$ equation $d = 0$ divide the half-plane
$(\xi, \varphi^{2})$, on three regions that will be denoted as $A$,
$B$ and $C$, while for $\eta = -1$ there are only two regions $A$ and
$B$ (figure 3a,b). The behavior of the model depends on the region
where the point $(\xi, \varphi^{2}_{0} )$ is situated.

\medskip

\begin{center}
 Fig. 3.  \\
The set $d = 0$ for $\eta = 1$ (3a) and $\eta = -1$ (3b) in
4-dimensions.
\end{center}

\medskip

Numerical investigation of equations (31)-(33) shows that for the
closed ($k = 1$) and flat ($k = 0$) cosmological models only singular
solutions exist for any initial conditions. For the open models
($k = -1$) if the pair $(\xi, \varphi^{2}_{0} )$ defines the
point in the region $B$ (both for $\eta = 1$ and $\eta = -1$) or $C$
(for $\eta = 1$) than only singular solutions of the equations
(37)-(39) exist. If the pair $(\xi, \varphi^{2}_{0} )$ defines the
point in the region $A$ then solutions may be both regular and
singular. The numerical investigation does not permit to find the
exact conditions of regularity, but it shows that both regular and
singular solutions are stable against finite perturbations of the
initial conditions. The typical qualitative behavior of the universe
scale factor $a(t)$, Weyl field $\omega$ and the matter scalar field
$\varphi$ are shown at figure 4a-c.

\begin{center}
 Fig. 4.  \\
 The qualitative behavior of the universe scale factor $a(t)$ (4a),
Weyl field $\omega$ (4b) and the matter scalar field $\varphi$ (4c)
in the open four-dimensional Weyl-integrable space-time model.
\end{center}

The universe scale factor $a(t)$ in the typical nonsingular
solution evolves from infinity at $t=-\infty$ to its minimal value
$a_{0} = a(0)$ and then grows to infinity at $t\rightarrow \infty$
(figure 4a). Both scalar fields, the Weyl field $\omega$ and the
field $\varphi$ evolves between two limiting values: from
$\omega_{-} = \lim_{t \rightarrow -\infty} \omega(t)$ and
$\varphi_{-} = \lim_{t \rightarrow -\infty} \varphi$ to
$\omega_{+} = \lim_{t \rightarrow \infty} \omega(t)$ and
$\varphi_{+} = \lim_{t \rightarrow \infty} \varphi(t)$. The difference
in the evolution of these fields is that the field $\omega$ evolves
monotonously (figure 4b) while the field $\varphi$ near $t=0$ (i. e.
near the minimum of $a(t)$) may have several intermediate extrema
with one absolute maximum if $\eta = 1$ (figure 4c) or absolute
minimum if $\eta = -1$. As $\varphi(t)$ for big $|t|$
tends asymptotically to constants, the model evolves asymptotically
as an empty Weyl cosmological model that is considered in section
2.1. It is necessary to note also that the evolution of the universe
scale factor $a(t)$ has a small time-asymmetry in comparison with the
case of the empty space. This asymmetry is a result of non
symmetrical evolution of the matter field $\varphi$ because the field
equations (37)-(39) contain both $\varphi$ and $\dot{\varphi}$.

{\it 3.2. 5-dimensional models}. In 5-dimensional case equations
(33)-(35) after simplification become
\begin{equation}         
 3\left\{ \frac{\dot{a}}{a} \left(\frac{\dot{a}}{a} +
   \frac{\dot{s}}{s} \right) +
   \frac{k}{a^{2}} \right\} \left\{ 1 + \frac{\varphi^{2}}{8}
\right\} + \frac{\varphi \dot{\varphi}}{4} \left( \frac{3\dot{a}}{a}+
   \frac{\dot{s}}{s} \right) + \frac{\eta \dot{\varphi}^{2}}{2} -
   \frac{\varphi \dot{\varphi} \dot{\omega}}{2} +
   \frac{\dot{\omega}^{2}}{2} \left( \xi - 3 -
    \frac{3\varphi^{2}}{8} \right) = 0,
\end{equation}
\begin{equation}              
 \left\{ \frac{\ddot{a}}{a} - \frac{\ddot{s}}{s} +
 2\frac{\dot{a}}{a} \left( \frac{\dot{a}}{a} -
 \frac{\dot{s}}{s} \right) + \frac{2k}{a^{2}} \right\} \left\{ 1 +
\frac{\varphi^{2}}{8} \right\} + \frac{\varphi \dot{\varphi}}{4}
\left(\frac{\dot{a}}{a} - \frac{\dot{s}}{s} \right) = 0,
\end{equation}
\begin{equation}              
  3 \left\{ 1 + \frac{\varphi^{2}}{8} \right\} \left\{
  \frac{\ddot{a}}{a} + \frac{\dot{a}}{a} \left( 2\frac{\dot{a}}{a} +
  \frac{\dot{s}}{s} \right) + \frac{2k}{a^{2}} \right\} +
  \frac{\varphi \ddot{\varphi}}{4} +
  \frac{\varphi \dot{\varphi}}{2} \left( 3 \frac{\dot{a}}{a} +
\frac{\dot{s}}{2s} + \frac{\dot{\varphi}}{2\varphi} \right) = 0,
\end{equation}
\begin{equation}              
  \ddot{\omega} \left( \frac{3\varphi^{2}}{4} - 2\xi + 6 \right) -
\varphi \ddot{\varphi} + \dot{\omega} \left( 3\frac{\dot{a}}{a} +
   \frac{\dot{s}}{s} \right) \left( 3\frac{\varphi^{2}}{4} - 2\xi +
   6 \right) - \varphi\dot{\varphi} \left( 3\frac{\dot{a}}{a} +
\frac{\dot{s}}{s} + \frac{\dot{\varphi}}{\varphi} \right) = 0,
\end{equation}
and
\begin{equation}         
\frac{\varphi}{2} \left( \frac{3\ddot{a}}{a} +
\frac{\ddot{s}}{s} \right) + \eta \ddot{\varphi} -
\varphi \ddot{\omega} +
\varphi \dot{\omega} \left( \frac{3\dot{\omega}}{4} -
3 \frac{\dot{a}}{a} - \frac{\dot{s}}{s} \right) + \eta \dot{\varphi}
\left( 3\frac{\dot{a}}{a} + \frac{\dot{s}}{s} \right) +
\frac{3\varphi \dot{a}}{2a} \left( \frac{\dot{a}}{a}+
\frac{\dot{s}}{s} \right) + \frac{3k \varphi}{2a^{2}} = 0.
\end{equation}

The equation (40) is the constraint that must be satisfied by
the initial conditions and the equations (41)-(44) are the dynamical.
The determinant of the matrix of the coefficients before
$\ddot{a}/a$, $\ddot{s}/s$, $\ddot{\omega}$ and $\ddot{\varphi}$ in
the dynamical equations (41)-(44) is equal to

\centerline{
$ d = \left( \frac{9}{256} \eta - \frac{3}{32} \right) \varphi^{6} +
\left( \frac{27}{32} \eta + \frac{1}{8} \xi - \frac{3}{32} \eta \xi -
\frac{3}{2} \right) \varphi^{4} + \left( \frac{27}{4}\eta + \xi -
\frac{3}{2}\eta\xi - 6 \right)\varphi^{2} + 18\eta - 6\eta\xi $.}

The qualitative features of function $d(\xi, \eta, \varphi)$ are
the same as in 4-dimensional case: for $\eta = 1$ equation $d = 0$
divides the half-plane $(\xi, \varphi^{2}>0)$ on three regions that
are denoted as $A$, $B$ and $C$, while for $\eta = -1$ there are only
two regions $A$ and $B$ (figure 5a,b). The behavior of the model
depends on the region where the point $(\xi, \varphi^{2}_{0} )$ is
situated.

\medskip

\begin{center}
 Fig. 5. \\
 The sets $d = 0$ for $\eta = 1$ (5a) and $\eta = -1$ (5b) in
5-dimensions.
\end{center}

\medskip

Numerical investigation of equations (40)-(44) shows that as well as
in the previous 4-dimensional case only singular solutions exist at
any initial conditions for the closed ($k = 1$) and flat ($k = 0$)
cosmological models. For the open models ($k = -1$) if the pair
$(\xi, \varphi^{2}_{0} )$ defines the point in the region $B$ (both
for $\eta = 1$ and $\eta = -1$) or $C$ (for $\eta = 1$) than only
singular solutions of the equations (40)-(44) exist, while if the
pair $(\xi, \varphi^{2}_{0})$ defines the point in the region $A$
than the solution may be both regular and singular. The regularity of
solutions depends on the constants of integration that may be
considered as the initial conditions at $t = 0$. It was
found that the regularity of solutions depends mainly on the signs of
$\dot{s}(0)$, $\dot{\omega}(0)$ and $\dot{\varphi}(0)$.
Their possible combinations that give nonsingular solutions of
equations (41)-(44) are represented in table 1. The last column of
this table shows the general direction of the interior space
evolution by means of the signs of the difference $\Delta = s_{+} -
s_{-}$, where $s_{\pm}=\lim_{t \rightarrow \pm \infty} s(t)$.

\begin{center}
 Table 1.\\
 Conditions of the solutions regularity and the direction of $s(t)$
evolution

\bigskip
\begin{tabular}{|c|c|c|c|} \hline
sign$\dot{s}(0)$ & sign$\dot{\omega}(0)$ & sign$\dot{\varphi}(0)$ &
sign($s_{+} - s_{-}$) \\   \hline
  -1   &     -1   &    0   &   -1    \\   \hline
  -1   &     +1   &    0   &   -1    \\   \hline
  -1   &     -1   &   +1   &   -1    \\   \hline
  -1   &     +1   &   -1   &   -1    \\   \hline
   0   &     +1   &    0   &   -1    \\   \hline
   0   &     -1   &    0   &   +1    \\   \hline
  +1   &     +1   &   -1   &   +1    \\   \hline
  +1   &     -1   &    0   &   +1    \\   \hline
  +1   &     +1   &    0   &   +1    \\   \hline
  +1   &     -1   &   +1   &   +1    \\   \hline
\end{tabular}
\bigskip

\end{center}

 The typical behavior of the nonsingular solution of the equations
(41)-(44) for $\eta = 1$ is shown at the figures 6a-d for
the case $\Delta \leq 0$, i. e. for the contracting interior
space.

\begin{center}

Fig. 6.  \\
 The qualitative behavior of the scale factors $a(t)$, (4a) and
$s(t)$ (4b), the matter scalar field $\varphi(t)$ (4c) and Weyl field
$\omega(t)$ (6d) in the nonsingular 5-dimensional Weyl-integrable
space-time model.

\end{center}

In general nonsingular solution the radius of the universe changes
monotonously from infinity at $t = - \infty$ to minimal value $a_{0}$
and then grows to infinity (figure 6a), while the radius of the
internal space starts from $s_{-}=\lim_{t \rightarrow -\infty}
s(t)$, passes through several (one or two) intermediate extrema, that
are situated near minimum of $a(t)$ and may be absent in some cases,
and then changes to $s_{+}=\lim_{t \rightarrow \infty} s(t)$
(figure 6b). Note that $s_{+}$ and $s_{-}$ may be of the same or
different order. The field $\varphi$ evolves analogously to
4-dimensional case (figure 6c). Note that the extremal points of
the functions $a(t)$, $s(t)$ and $\varphi (t)$ do not coincide with
each other in general case and the function $a(t)$ is time
asymmetrical especially near its minimum. Finally the Weyl field
$\omega$ changes monotonously between two limiting values:
$\omega_{-}=\lim_{t \rightarrow - \infty} \omega(t)$ and
$\omega_{+}=\lim_{t \rightarrow  \infty} \omega(t)$ (figure 6d). In
the case $\eta = -1$ the model evolves as above but the extremal
points of the field $\varphi$ change type: minimum become maximum and
vice versa.

\section{Concluding remarks}

We have considered the qualitative evolution of multidimensional
cosmological models based on the integrable Weyl geometry both in
vacuum space-time and in the presence of nonminimal scalar field. The
existence of nonsingular solutions of field equations for open
cosmological models that realized the dimensional reduction scenario
was demonstrated. It was shown that in multidimensional case the
evolution of the scale factor of the universe $a(t)$ becomes
time-asymmetric unlike the four-dimensional case. We have shown also
that all nonsingular cosmological models considered above have some
common features. In particular the evolution of the universe scale
factor (radius) $a(t)$ for big $|t|$ is asymptotically linear.
Further in all nonsingular models Weyl scalar field $\omega (t)$ as
well as the matter field $\varphi (t)$ in the models with nonminimal
coupling tend asymptotically to constants. So the models tend to the
pure Einsteinian models of the corresponding dimensions and the
change of the collapse era into expansion one may be considered as a
cosmological phase transition induced by the transition of scalar
fields $\omega (t)$ and $\varphi (t)$ from one stationary state
$\omega = \omega_{-}$ and $\varphi = \varphi_{-}$ into another
stationary state $\omega = \omega_{+}$ and $\varphi = \varphi_{+}$.
At the late stages of the universe evolution the fields $\omega (t)$
and $\varphi (t)$ are unobservable.

There are several qualitative differences between the vacuum models
and the models with nonminimal scalar field. First of all in vacuum
models the existence of cosmological singularity depends only on the
parameters of the theory while in the case of nonminimal scalar
field it depends on the initial conditions also. Secondly, in the
models with nonminimal scalar field the evolution of the internal
space scale factor $s(t)$ may be nonmonotonous. In the typical
scenario one of the limiting values of $s(t)$ at $t = \pm \infty$
is much smaller than another but in several models both limiting
values of internal radius $s(t)$ may be arbitrary small and it
become finite only near minimum of the universe scale factor $a(t)$.

We have discussed here only the models with the one- or two-
dimensional interior space because if interior space has
dimension $d \geq 3$ and direct product topology of torus on several
spheres then the models have the same qualitative features as
considered above. In particcular, the nonsingular solutions exist
only for toroidal interior space topology.

The models considered above show that the real geometrical structure
of space-time may have a non-Riemaniann nature but the universe may
evolve in such a way that its non-Riemaniann nature is essential only
near $t = 0$ and become unobservable at late stages of the evolution.
Therefore, the consideration of generalized geometrical structures
in multidimensional cosmology may be of a considerable interest. In
particular, the models considered above may be generalized in the
following manner. First of all, both Weyl scalar field $\omega (t)$
and matter field $\varphi (t)$ may be massive and have nonlinear
potential. Secondly, the possible influence of the cosmological term
$\Lambda$ must be considered also. At last, the term
$R\varphi^{2}/2(n-1)$ in the action integral (25) may have negative
sign. One may suppose that in this case nonsingular solutions of the
field equations may be obtained not only for open models, but for
closed and flat models also. These possibilities will be considered
elsewhere.

\section{Acknowledgments}

This work was supported in part by the Russian Ministry of Science.

\end{document}